\documentclass[twocolumn,english,aps,prl,showpacs,superscriptaddress]{revtex4-1}
\usepackage[T1]{fontenc}
\usepackage[latin9]{inputenc}
\usepackage{xcolor}
\usepackage{pdfcolmk}
\usepackage{amsmath}
\usepackage{amssymb}
\usepackage{graphicx}
\PassOptionsToPackage{normalem}{ulem}
\usepackage{ulem}

\makeatletter

\providecolor{lyxadded}{rgb}{0,0,1}
\providecolor{lyxdeleted}{rgb}{1,0,0}

\DeclareRobustCommand{\lyxsout}[1]{\ifx\\#1\else\sout{#1}\fi}

\usepackage{babel}

\makeatother

\usepackage{babel}
\begin{document}
\title{Multiport Atom Interferometry for Inertial Sensing}
\author{Dimitry Yankelev}
\email[These authors contributed equally to this work. Electronic addresses: dimitry.yankelev@weizmann.ac.il, chen.avinadav@weizmann.ac.il.]{}

\affiliation{Department of Physics of Complex Systems, Weizmann Institute of Science,
Rehovot 7610001, Israel}
\affiliation{Rafael Ltd, Haifa 3102102, Israel}
\author{Chen Avinadav}
\email[These authors contributed equally to this work. Electronic addresses: dimitry.yankelev@weizmann.ac.il, chen.avinadav@weizmann.ac.il.]{}

\affiliation{Department of Physics of Complex Systems, Weizmann Institute of Science,
Rehovot 7610001, Israel}
\affiliation{Rafael Ltd, Haifa 3102102, Israel}
\author{Nir Davidson}
\affiliation{Department of Physics of Complex Systems, Weizmann Institute of Science,
Rehovot 7610001, Israel}
\author{Ofer Firstenberg}
\affiliation{Department of Physics of Complex Systems, Weizmann Institute of Science,
Rehovot 7610001, Israel}
\begin{abstract}
We present new techniques for inertial-sensing atom interferometers
which produce multiple phase measurements per experimental cycle.
With these techniques, we realize two types of multiport measurements,
namely quadrature phase detection and real-time systematic phase cancellation,
which address challenges in operating high-sensitivity cold-atom sensors
in mobile and field applications. We confirm experimentally the increase
in sensitivity due to quadrature phase detection in the presence of
large phase uncertainty, and demonstrate suppression of systematic
phases on a single shot basis.
\end{abstract}
\maketitle
Cold atom interferometers have demonstrated extremely high sensitivity
as inertial sensors measuring gravity \cite{Kasevich_1991,Peters1999,Peters2001},
gravity gradients \cite{Snadden1998,McGuirk2002,Sorrentino2012,Sorrentino2014,Biedermann2015},
accelerations and rotations \cite{Barrett2014,Canuel2006,Stockton2011,Wu2014,Savoie2018,Gustavson1997,Dickerson2013,Hoth2016,Chen2018}.
In addition to precision measurements of physical constants \cite{Weiss1993,Fixler2007,Bouchendira2011,Rosi2014,Parker2018},
tests of general relativity \cite{Dimopoulos2007,Chung2009,Mueller2010,Hohensee2011,Aguilera2014,Zhou2015a},
searches for dark energy \cite{Hamilton2015,Jaffe2017}, and gravitational
wave detection \cite{Graham2013,Canuel2018}, they are promising candidates
as on-board inertial measurement units \cite{Canuel2006,Geiger2011,Wu2017,Cheiney2018}
and as mobile gravimeters for geodesic studies or subterranean exploration
\cite{Wu2009,Bongs2014,Freier2016,Bidel2018,Menoret2018,Zhu2018}.
These prospects provide strong motivation for improving the robustness
of atom interferometers while maintaining high phase sensitivity and
accuracy under field conditions, such as strong vibrations and drifts
in the thermal and magnetic environment.

Atom interferometers typically yield a single phase measurement per
shot. This may limit their performance in challenging conditions:
first, phase sensitivity is maximal in the linear regime near mid-fringe,
which requires locking the phase from shot to shot \cite{Menoret2018}.
This is difficult to maintain when the inertial signal changes on
short timescales, such as in mobile applications, or when vibrations
introduce large uncontrolled phase variation. Real-time correction
using classical sensors can be applied to return the interferometer
to mid-fringe \cite{Lautier2014,Menoret2018}, but effects such as
delay in their response \cite{Merlet2009} ultimately limit their
effectiveness for strong vibrations. Second, many systematic phase
shifts are typically canceled with the ``$k$-reversal'' technique,
using sequential measurements with opposite wave vectors \cite{McGuirk2002,Louchet-Chauvet2011}.
However, it is not effective against fast variations in systematic
effects, which may arise in field applications, and it inherently
reduces the bandwidth and sensitivity per $\sqrt{\text{Hz}}$ of the
interferometer.

In this Letter, we introduce two new schemes which extend inertial-sensing
atom interferometers to yield multiple phase signals in each experiment,
increasing its information bandwidth and improving its performance.
One scheme utilizes a composite beam-splitter, which replaces the
final beam-splitter in typical atom interferometers, and the other
is based on operating dual concurrent interferometers on a single
atomic ensemble, with independent control of their phases. Using these
schemes, we demonstrate two measurement approaches which address challenges
of deployable atom interferometers. The first approach realizes quadrature
phase detection, maintaining peak sensitivity at all phases and avoiding
low-sensitivity shots away from mid-fringe. The second approach uses
concurrent $k$-reversal interferometry for real-time cancellation
of systematic phase shifts.

\begin{figure*}
\begin{centering}
\includegraphics[clip,width=2\columnwidth]{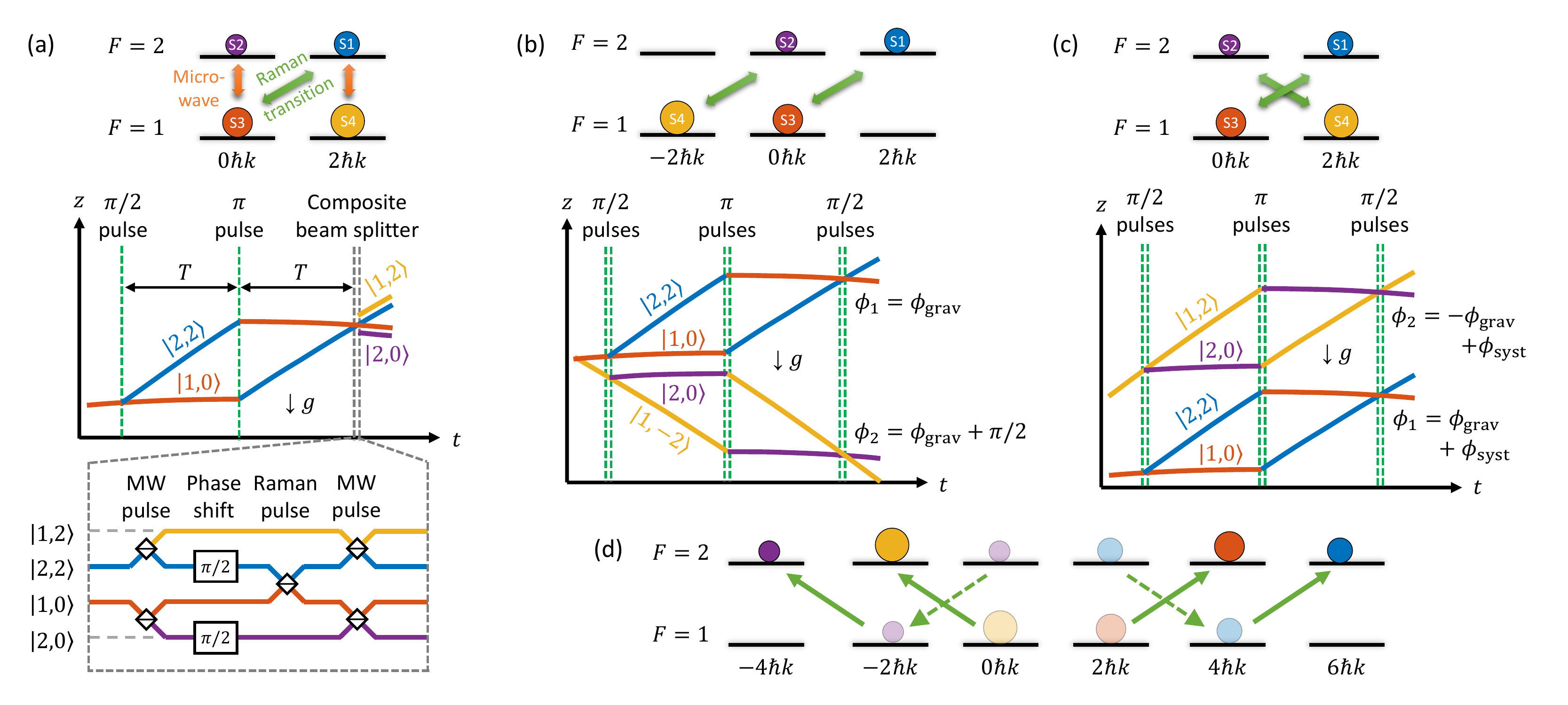}
\par\end{centering}
\centering{}\caption{Multiport interferometry and detection. (a) Quadrature detection with
composite beam-splitter. The composite operation (inset) consists
of a $\pi/3$ microwave pulse, a $\pi/2$ differential phase shift,
a narrowband $0.6\pi$ Raman pulse, and a final $\pi/2$ microwave
pulse, resulting in $\pi/2$ phase difference between the interference
in states $\left|1,0\right\rangle ,\left|2,2\right\rangle $ and $\left|1,2\right\rangle ,\left|2,0\right\rangle $.
$S_{i}$ represent fractional populations in the four relevant states,
in arbitrary order. (b) Quadrature detection with concurrent interferometry.
Atoms are initialized in an equal mixture of $\left|1,0\right\rangle $
and $\left|1,-2\right\rangle $, which undergo simultaneous Raman
transitions. The relative phase between the Raman fields is changed
by $\pi/2$ before the final beam-splitter pulse. (c) Real-time $k$-reversal
with concurrent interferometry. Initial states are $\left|1,0\right\rangle $
and $\left|1,2\right\rangle $, and the two Raman transitions are
tuned to impart opposite momentum kicks. (d) Example of multiport
detection scheme. Dashed arrows represent shelving of $F=2$ populations
in $F=1$ followed by blow-away, solid arrows represent sequential
promotions to $F=2$, each followed by resonant fluorescent detection
and blow-away. This example corresponds to detection of interferometers
(a) and (c); detection of (b) differs only in some shelving and promotion
operations.\label{fig:schemes}}
\end{figure*}

Atom interferometers based on stimulated two-photon Raman transitions
\cite{Kasevich1991a,Moler1992} use two atomic states separated by
momentum and internal state, e.g. $\left|1,0\right\rangle \equiv\left|F=1,p=0\hbar k\right\rangle $
and $\left|2,2\right\rangle \equiv\left|F=2,p=2\hbar k\right\rangle $,
where $k$ is the wave vector of the Raman lasers. In the following
we consider the Mach-Zehnder sequence with three Raman pulses $\pi/2$-$\pi$-$\pi/2$
equally separated by time $T$ \cite{Kasevich_1991}. The first beam-splitter
pulse generates a coherent superposition of $\left|1,0\right\rangle $
and $\left|2,2\right\rangle $ which then spatially drift apart, followed
by a mirror pulse which redirects them back towards each other. The
final beam-splitter mixes the states coherently and produces interference
in their populations, which constitute the output ports of the interferometer.
Its phase in this sequence is $\phi=\mathbf{k}_{\text{eff}}\cdot\mathbf{a}T^{2}$
where $\mathbf{k}_{\text{eff}}\cong2\mathbf{k}$ is the effective
wave vector of the Raman interaction and $\mathbf{a}$ the relative
acceleration between the atoms and the mirror reflecting the Raman
beams. For a gravimeter this simplifies to $\phi_{\text{grav}}=k_{\text{eff}}gT^{2}$.
Our multiport interferometry schemes exploit the momentum degree of
freedom to produce additional output signals with different phases
and extract more information from each experiment cycle.

In the first scheme, a composite beam-splitter replaces the final
beam-splitter pulse with a combination of microwave rotations, state-dependent
light shifts, and narrowband Raman transition. This operation couples
two additional states, $\left|1,2\right\rangle $ and $\left|2,0\right\rangle $
{[}Fig. \ref{fig:schemes}(a){]}, whose interference contrast and
phase is determined by the pulse parameters of the composite beam-splitter.
We numerically found such values, depicted in Fig. \ref{fig:schemes}(a),
which produce two pairs of interference signals in phase quadrature.

Conversely, the concurrent interferometry scheme relies on initializing
the atomic ensemble in two different momentum states of the same ground
state and applying the interferometric pulse-sequence to both states
simultaneously, resulting in four output ports. By independently controlling
the phase and frequency of the narrowband Raman pulses which address
each momentum state, we realize different types of multiport measurements:
quadrature detection, where the phase of the Raman pulses differs
by $\pi/2$ on the final beam-splitter pulse {[}Fig. \ref{fig:schemes}(b){]};
and $k$-reversal, where the frequency of the Raman pulses is chirped
in opposite directions to address opposite momentum transitions {[}Fig.
\ref{fig:schemes}(c){]}, changing the sign of $\phi_{\text{grav}}$
but not the sign of many systematic phases, denoted collectively as
$\phi_{\text{syst}}$ \cite{Louchet-Chauvet2011}.

Measuring the multiple output ports in either scheme requires velocity-
and state-dependent detection. Constrained by small spatial separation
of the output states, we use a combination of velocity-selective Raman
pulses and state-selective resonant detection. Following the interferometry
sequence, $F=2$ populations are first transferred to empty momentum
states in $F=1$. Then, sequentially, each momentum state is transferred
to $F=2$, where it is counted and blown away by a resonant light
pulse {[}Fig. \ref{fig:schemes}(d){]}.

Both multiport schemes are compatible with many Raman atom interferometers,
including different pulse sequences such as $\pi/2$-$\pi$-$\pi$-$\pi/2$
for rotation sensing \cite{Canuel2006,Stockton2011,Wu2014,Savoie2018}
and large momentum transfer interferometers \cite{McGuirk2000,Kotru2015}.
Interferometers operating in micro-gravity \cite{Geiger2011}, where
atoms are scattered into multiple momentum states due to vanishing
Doppler shift \cite{Malossi2010}, may also benefit from multiport
interferometry and detection. Concurrent interferometry should also
be compatible with zero-dead-time and interleaved operation \cite{Savoie2018}.
Comparing the two multiport schemes, the composite beam-splitter has
the advantage that all output ports result from a single interferometer
loop and fundamentally measure the same inertial phase, whereas the
concurrent interferometer does not require stabilized light-shift
pulses or microwave rotations, and allows for more versatile multiport
configurations such as $k$-reversal. Previous applications of concurrent
atom interferometers were limited to gravity gradiometers, which are
inherently insensitive to inertial forces, or to Ramsey-Bordé interferometers
\cite{Chiow2009,Estey2015,Zhong2019} in measurements of the fine-structure
constant. Quadrature detection was previously realized using two different
atomic species to operate two interferometers \cite{Bonnin2018},
which adds experimental complexity and is susceptible to uncommon
noise sources of the two species.

Our apparatus prepares an ensemble of $\sim10^{8}$ $^{87}\text{Rb}$
atoms at $\sim6\,\mu\text{K}$ and launches it upwards at velocities
of up to $1.5\,\text{m/s}$ using moving optical molasses. Retro-reflected
Raman beams with $\sigma^{+}/\sigma^{+}$ polarizations traverse the
system vertically and interact with the atoms along their trajectory
for initialization, interferometry, and multiport detection. The Raman
mirror is placed on a vibration isolation platform whose residual
vibrations are measured using a classical accelerometer. The Raman
beams originate from a $780\textrm{ nm}$ DBR laser, locked 700 MHz
below the $F=2\rightarrow F'=1$ transition and passing through a
fiber electro-optic modulator (EOM) and a semiconductor optical amplifier.
The Raman beams have $40\text{ mm}$ $1/e^{2}$ diameter and a total
power of 200 mW in all sidebands. For simultaneous addressing of two
velocity classes in concurrent interferometry, the EOM microwave signal
contains two spectral components independently generated by a direct
digital synthesizer and up-shifted to approximately $6.834\,\text{GHz}$
using a single-sideband mixer. A horizontal beam resonant with $F=2\rightarrow F'=3$
is used for blow-away and fluorescence detection.

The atoms initially populate all Zeeman sublevels of the $F=2$ manifold
and have a thermal momentum distribution with $\sim9\hbar k$ FWHM.
Narrowband Raman pulses select atoms from $m_{F}=0$ with a momentum
spread of $0.7\hbar k$ FWHM. For the composite beam-splitter, the
selection is centered on the mean ensemble momentum $\bar{p}$, while
for concurrent interferometry, we select two groups at $\bar{p}\pm\hbar k$.
The latter is possible due to the wide thermal distribution; a colder
ensemble could be initialized in two different momentum or hyperfine
states by microwave, Raman, or Bragg transitions. The interferometer
pulses have a bandwidth of $1.1\hbar k$ FWHM, which allows addressing
the velocity-selected states with high fidelity and minimal cross-talk
between states separated by $2\hbar k$.

The state-dependent phase-shift for the composite beam-splitter is
implemented with a laser beam detuned 1 GHz below the $F=2\rightarrow F'=3$
transition, inducing a differential light shift between $F=1$ and
$F=2$. The pulse duration and power are first calibrated using standard
Raman-interferometry to create the necessary $\pi/2$ phase shift.
This pulse has similar duration, detuning and intensity as the interferometer
Raman pulses and thus additional scattering is insignificant.

\begin{figure}
\begin{centering}
\includegraphics[clip,width=1\columnwidth]{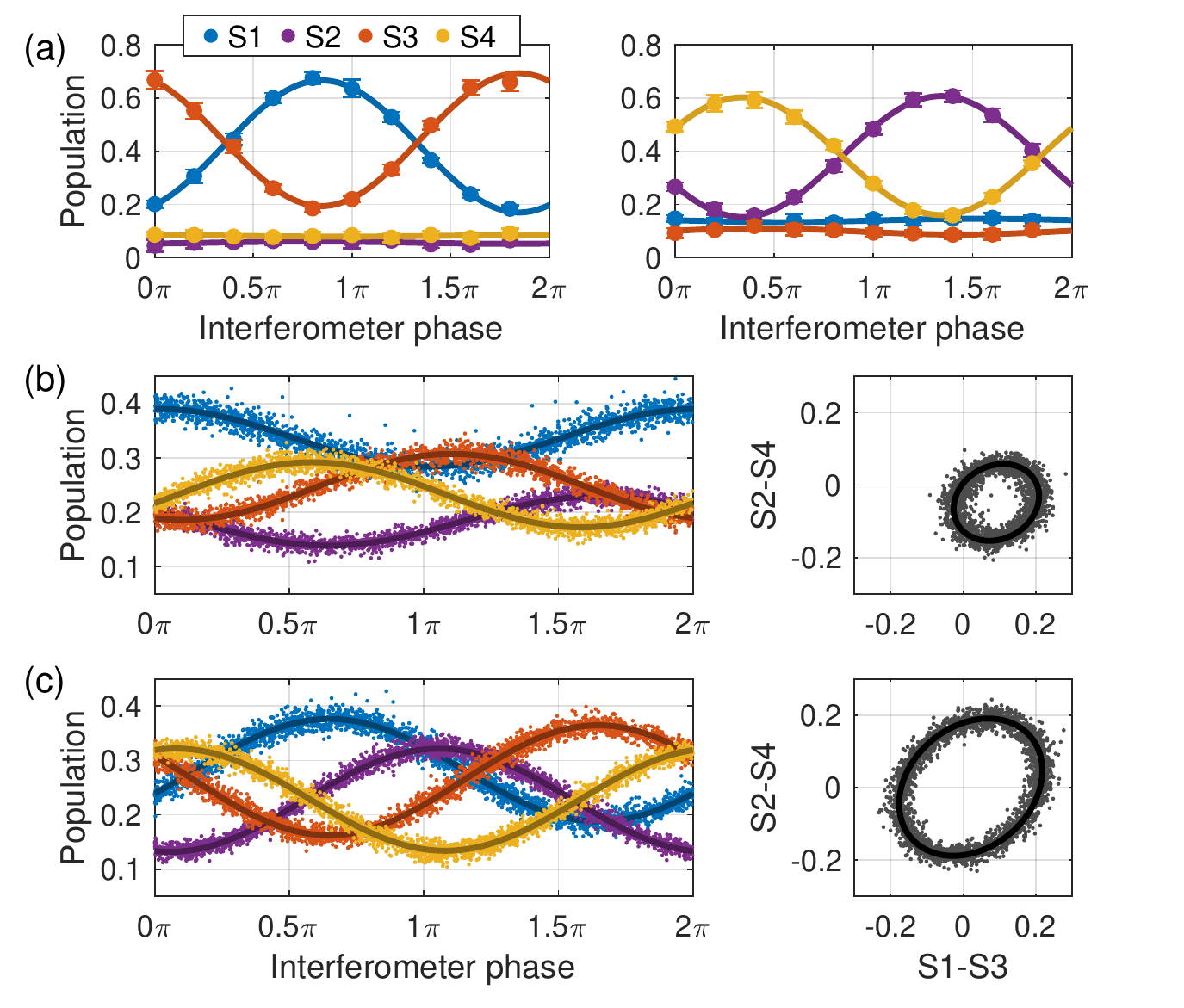}
\par\end{centering}
\centering{}\caption{(a) Characterization of multiport detection. Four output ports measured
following a standard interferometer sequence, when populating only
the initial states $\left|1,0\right\rangle $ (left) or $\left|1,-2\right\rangle $
(right). The cross-talks, measured by the contrast ratios between
the initially unpopulated and populated states, are $(1.4\pm0.8)\%$
and $(3.9\pm1.2)\%$ respectively. Solid lines are least-squares fringe
fits. (b) Four output signals in the composite beam-splitter scheme
(left) and their representation as an ellipse in a Lissajous plot
(right). The contrast ratio between the quadrature fringes is 1.11
and their relative phase $0.45\pi$. The Raman laser phase was scanned
randomly over $2\pi$. (c) Same as (b) for the concurrent interferometry
scheme. The contrast ratio here is 1.04 and the relative phase $0.42\pi$.
Mean contrast is 70\% higher than for the composite beam-splitter
due to the lack of additional microwave and light shift operations.
Data measured with $2T=2\text{ ms}$. \label{fig:quad-raw-data}}
\end{figure}

We characterize the multiport detection sequence by selecting atoms
in one momentum state, performing standard Mach-Zehnder interferometry,
and measuring the population of all four output ports. As Fig. \ref{fig:quad-raw-data}(a)
shows, high-contrast fringes are measured in the expected output states
with little cross-talk between the pairs, demonstrating the high fidelity
of our detection scheme. Such measurements can be used to calibrate
the detection process transformation and correct its imperfections.

\begin{figure}
\begin{centering}
\includegraphics[viewport=0bp 0bp 396bp 264bp,clip,width=1\columnwidth]{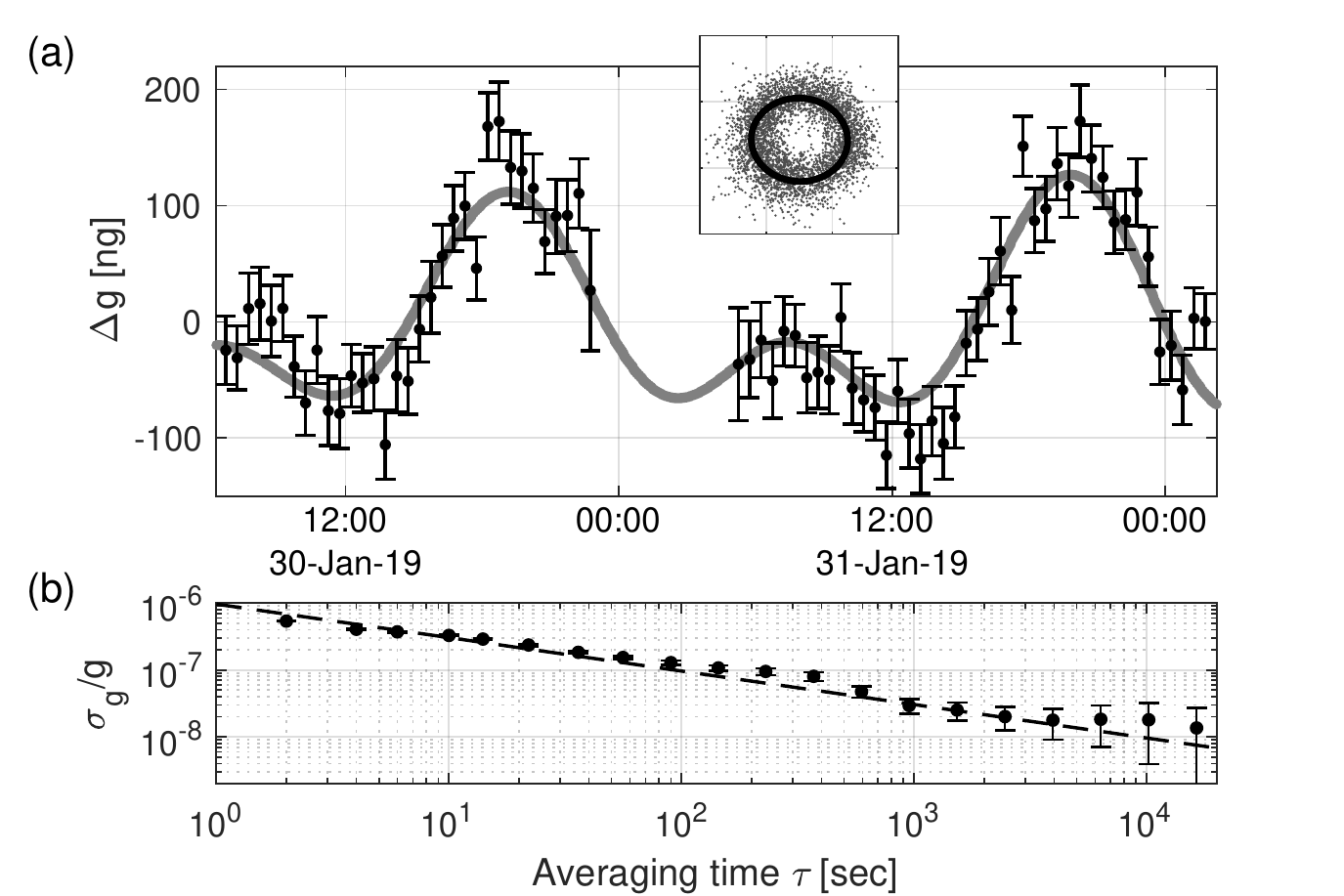}
\par\end{centering}
\centering{}\caption{Sensitivity and long-term stability of concurrent quadrature interferometry.
(a) Gravity variations measured over 44 hours with $2T=140\text{ ms}$.
Inset shows raw data over several hours. Time is in UTC, data points
are averaged over 30-minute bins, solid line is the tidal model calculated
at the measurement site (Haifa, Israel). (b) Allan deviation of gravity
measurements after subtracting the tidal model. Dashed line is a fit
to $\tau^{-1/2}$. \label{fig:quad-tides}}
\end{figure}

A demonstration of quadrature detection using both composite beam-splitter
and concurrent interferometry schemes is shown in Fig. \ref{fig:quad-raw-data}(b-c).
The four output signals in both cases are in nearly perfect quadrature.
We analyze quadrature measurements by taking the differential population
of each pair of signals and fitting an ellipse to their Lissajous
curve. Once the ellipse fit is known, we estimate the phase for each
individual measurement from the nearest point on the ellipse.

We test the performance of quadrature detection by realizing concurrent
interferometry with $2T=140\text{ ms}$ and measuring gravity variations
over 44 hours (Fig. \ref{fig:quad-tides}). The results follow the
expected changes due to gravity tides, and residuals from the tidal
model indicate a sensitivity of $0.96\,\mu\text{g}/\sqrt{\text{Hz}}$,
consistent with technical noise sources in our apparatus, and stability
of $13\text{ ng}$ at $1.6\times10^{4}\text{ sec}$. The measurements
presented in Fig. \ref{fig:quad-raw-data}(b-c) also maintained stability
for the duration of the run (6 and 15 hours respectively; data not
shown). This demonstrates the compatibility of our quadrature detection
method for high-sensitivity operation and its long-term stability.

\begin{figure}[t]
\begin{centering}
\includegraphics[viewport=0bp 10bp 396bp 396bp,clip,width=1\columnwidth]{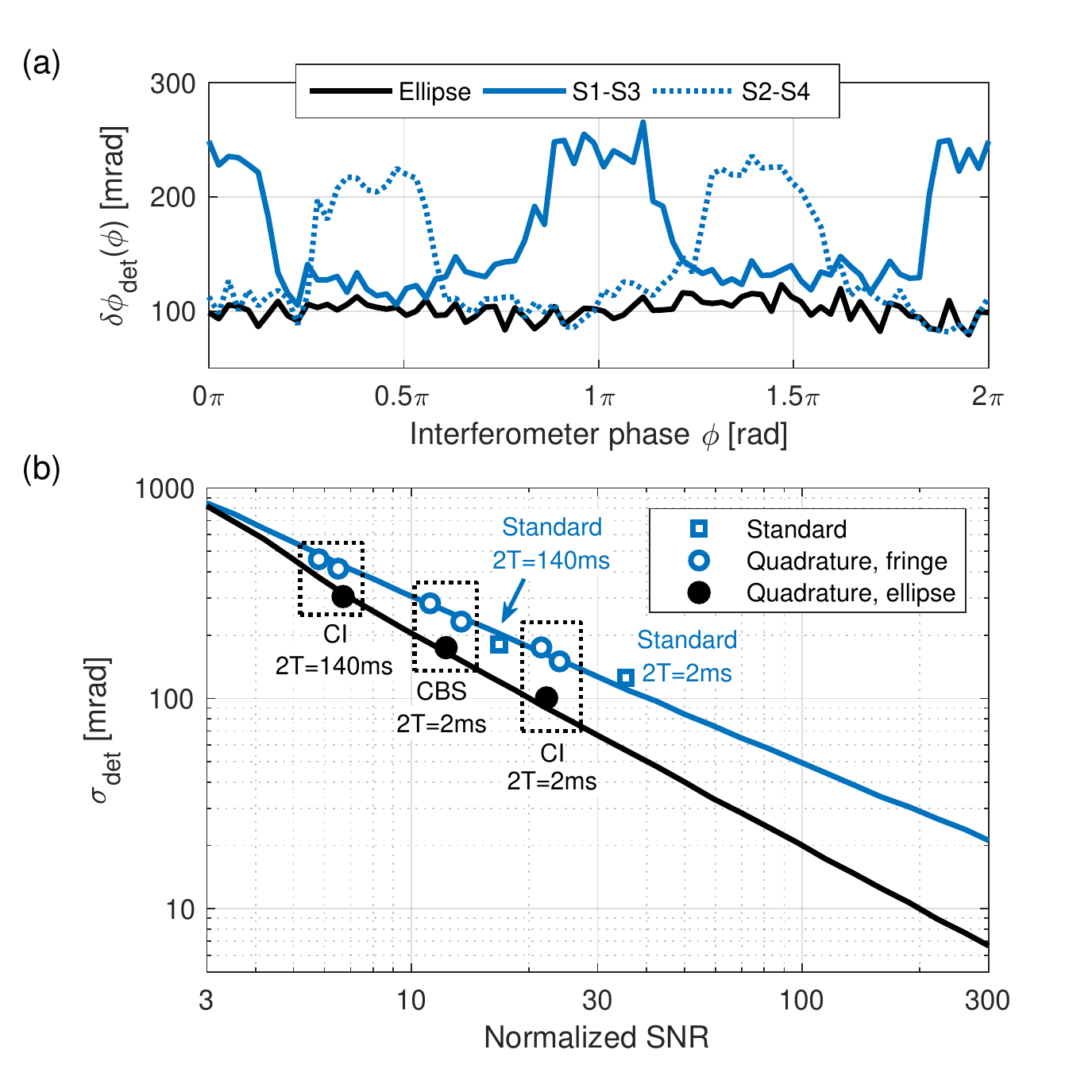}
\par\end{centering}
\centering{}\caption{Analysis of phase sensitivity. (a) Standard deviation of phase uncertainty
due to detection noise $\delta\phi_{\text{det}}$, as a function of
the interferometer phase $\phi$ for concurrent interferometry with
$2T=2\text{ ms}$, using ellipse analysis of all output ports (black)
or single-fringe analysis of each pair separately (blue). Results
extracted from 14,000 measurements. (b) Overall phase uncertainty
$\sigma_{\text{det}}$ as a function of normalized SNR for ellipse
analysis (black) and single-fringe analysis (blue). Markers represent
experimental results at several conditions, with the SNR estimated
from the measured data. Lines represent numerical simulations. ``Standard''
stands for regular two-port interferometry, CI is concurrent interferometry,
and CBS is composite beam-splitter. Cloud expansion at longer $T$
lowers the fidelity of the Raman pulses and consequently decreases
the SNR.\label{fig:phase-sens-analysis}}
\end{figure}

We now turn to a detailed analysis of the improved phase sensitivity
of quadrature versus standard fringe detection. We consider a single
interferometric signal, $S=A-\frac{C}{2}\cos(\phi+\delta\phi)+\delta S$,
where $A$ and $C$ are the fringe offset and contrast, $\phi$ the
interferometer phase, $\delta\phi\sim N\left(0,\sigma_{\phi}\right)$
represents laser phase noise or imperfect vibration correction \footnote{We assume Gaussian noise in the discussion for simplicity.},
and $\delta S\sim N\left(0,\sigma_{\text{S}}\right)$ is detection
noise with signal-to-noise ratio (SNR) of $C/\sigma_{\text{S}}$.
$\sigma_{\text{S}}$ and $\sigma_{\phi}$ can be estimated from the
measurement noise at fringe extrema and mid-fringe, respectively.
The total phase uncertainty is given by $\delta\phi_{\text{tot}}=\delta\phi_{\text{det}}+\delta\phi$,
with $\delta\phi_{\text{det}}$ the uncertainty from a single data
point due to finite SNR. $\delta\phi_{\text{det}}$ depends not only
on $\sigma_{\text{S}}$ but also on $\phi$ itself, with its variance
being minimal near mid-fringe and maximal near its extrema. Conversely,
in ellipse analysis of quadrature signals this term is independent
of $\phi$ and equals to the mid-fringe uncertainty of single-fringe
analysis with the same SNR. We demonstrate this experimentally in
Fig. \ref{fig:phase-sens-analysis}(a), clearly highlighting the advantage
of quadrature detection when the interferometer phase varies far from
mid-fringe.

To quantify the performance of quadrature detection, we estimate the
overall phase uncertainty due to SNR, denoted as $\sigma_{\text{det}}$,
from measurements where the phase varies randomly over $2\pi$. Assuming
independent detection and phase noises, we have $\sigma_{\text{det}}^{2}=\sigma_{\text{tot}}^{2}-\sigma_{\phi}^{2}$,
with $\sigma_{\text{tot}}^{2}$ the two-sample Allan variance of the
estimated phase. We present the results in Fig. \ref{fig:phase-sens-analysis}(b)
for the two quadrature schemes (concurrent interferometry and composite
beam-splitter) and for standard two-port interferometry, at various
SNR values and interrogation times $T$. We observe good agreement
with the numerical simulations of the two detection schemes. For additional
validation of the simulations, we also present the single-fringe analysis
of the quadrature measurements. Notably, quadrature detection outperforms
standard fringe detection at any given SNR. The relative sensitivity
gain increases with increasing SNR, e.g. from 1.5 to 2.5 for SNR values
from 10 to 100. Consequently, reduction in SNR due to implementation
of quadrature detection becomes less significant at higher SNR values.

\begin{figure}
\begin{centering}
\includegraphics[viewport=0bp 0bp 378bp 248bp,clip,width=1\columnwidth]{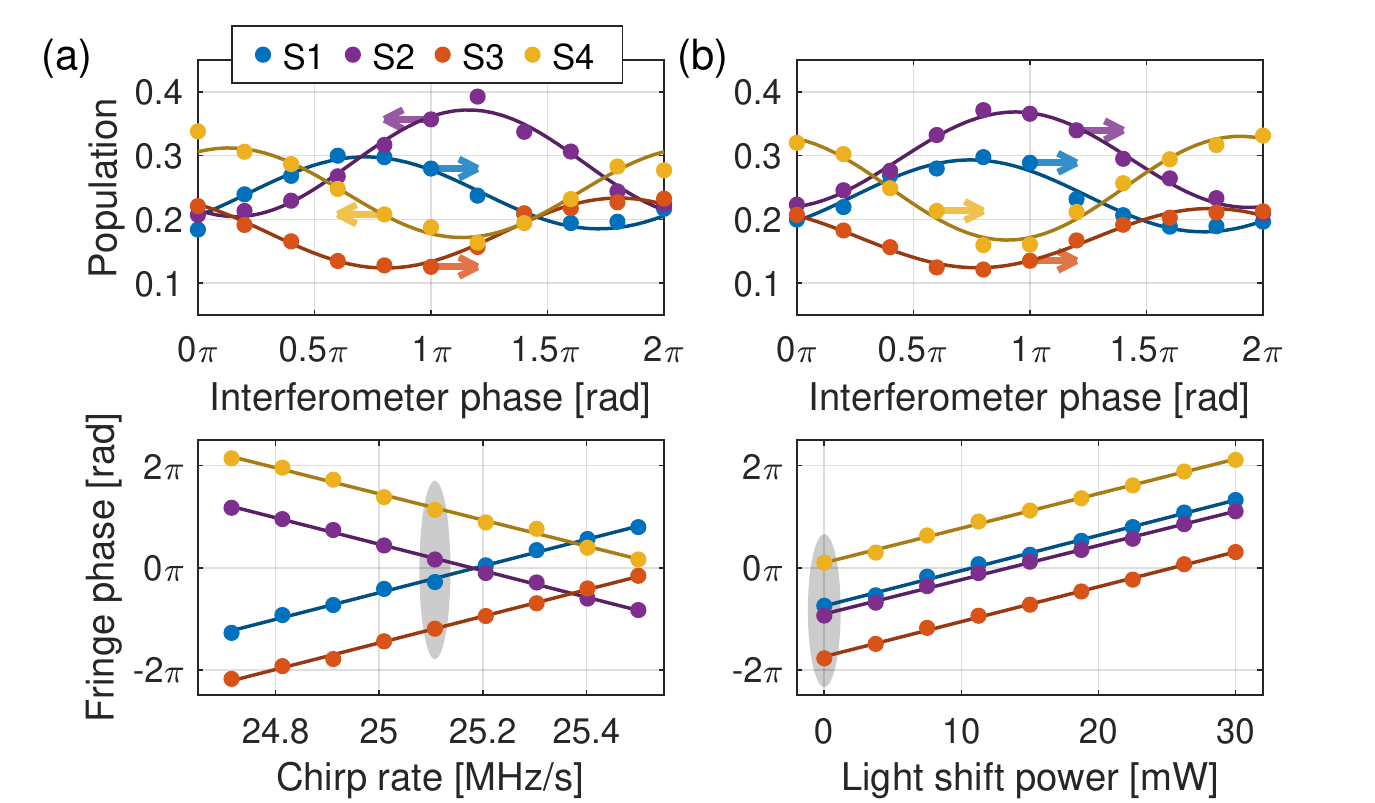}
\par\end{centering}
\centering{}\caption{Concurrent $k$-reversal interferometry. (a) Top: Raw output signals
measured for a specific value of the Raman chirp rate. Bottom: The
fringe phases of each output port as a function of chirp rate. As
the chirp rate changes, representing a change in the measured acceleration,
the two pairs of fringes shift in opposite directions and their relative
phase is linearly proportional to the applied chirp. Shaded points
in bottom figure correspond to the fringes shown above. (b) Same as
(a), but as a function of the light-shift pulse power, simulating
different systematic shifts, at a fixed chirp rate. In this case,
all fringes shift in the same direction and their relative phase remains
stable despite the changing systematic phase. Data measured with $2T=4\text{ ms}$.
\label{fig:k-reversal}}
\end{figure}

Finally, we present proof-of-principle demonstration of concurrent
interferometry for real-time systematic phase-shift cancellation using
$k$-reversal. In these experiments, we control the inertial phase
by changing the chirp rate of the Raman lasers \cite{Louchet-Chauvet2011},
and we simulate systematic phase shifts by adding a light-shift pulse
to the interferometer and varying its power. As expected, we find
the relative phase between pairs of opposite-$k$ fringes corresponds
to the inertial phase, whereas their common phase corresponds to the
systematic effect (Fig. \ref{fig:k-reversal}). We observe that even
for large systematic contributions of $2\pi$, we extract the same
inertial phase up to $33\text{ mrad}(1\sigma)$. This indicates a
lower bound for the suppression ratio of $190\text{:}1$, likely limited
by SNR in these measurements. Contrary to traditional sequential implementation
of the $k$-reversal technique, the method presented here allows evaluation
of both inertial and systematic phases from a single shot of the experiment
by measuring the multiple ports of the interferometer.

In conclusion, we introduce new techniques that bring multiport detection
capabilities to inertial-sensing atom interferometers. The two techniques,
a composite beam-splitter and concurrent interferometry, increase
the phase information measured per experimental cycle. We demonstrate
two applications of multiport interferometry: quadrature phase detection,
which maintains constant and high phase sensitivity and bandwidth
in the presence of large phase variations, and real-time $k$-reversal,
capable of measuring and canceling systematic effects even as they
change on fast timescales. Both applications are expected to increase
the performance and robustness of future quantum inertial sensors
operating in challenging field conditions, in the face of large and
uncontrolled phase variations and noisy environments. The methods
presented in this paper can be readily extended to additional output
ports, e.g. 6- or 8-port interferometers. For example, advanced sensors
based on dual-wavelength interferometers, e.g., $780$ and $795\textrm{ nm}$,
with dramatically increased dynamic range due to their different scale
factors, would require quadrature detection to avoid phase ambiguities.
\begin{acknowledgments}
We thank Yoav Erlich and Igal Levy for technical assistance. This
work was supported by the Pazy Foundation.
\end{acknowledgments}


%

\end{document}